\documentclass[11pt]{article}
\usepackage{graphicx}
\usepackage{amsmath}
\usepackage{amssymb}
\usepackage{caption2}
\begin{document}
\bibliographystyle{prsty}
\begin{center}
{\large {\bf \sc{  Analysis of mass difference of the $\pi$ and $\rho$ with Bethe-Salpeter equation }}} \\[2mm]
Zhi-Gang Wang \footnote{E-mail,wangzgyiti@yahoo.com.cn.  }    \\
$^{1}$ Department of Physics, North China Electric Power University,
Baoding 071003, P. R. China
\end{center}

\begin{abstract}
In this article, we take  into account the one-pion exchange force
besides   the one-gluon exchange force to study the mass difference
of the $\pi$ and $\rho$ mesons  with the Bethe-Salpeter equation.
After projecting the Bethe-Salpeter equation into  an simple form,
we can see explicitly that the bound energy $|E_\pi|\gg |E_\rho|$.
\end{abstract}



\section{Introduction}
The constituent quark models have given many successful descriptions
of the   hadron spectroscopy, the simple constituent quark mass plus
hyperfine spin-spin interaction model works well for the ground
state mesons \cite{Rosner81},
\begin{eqnarray}
 M_m&=&M_1+M_2+C\frac{\vec{\sigma}_1 \cdot \vec{\sigma}_2}{M_1M_2}\, ,
\end{eqnarray}
where the coefficient $C$ can be fitted phenomenologically. The
masses of the ground state pseudoscalar and vector mesons are
$M_\pi=140\,\rm{MeV}$, $M_K=494\,\rm{MeV}$, $M_\eta=548\,\rm{MeV}$,
$M_{\eta'}=958\,\rm{MeV}$, $M_\rho=775\,\rm{MeV}$,
$M_{\omega}=783\,\rm{MeV}$, $M_{K^*}=892\,\rm{MeV}$, and
$M_{\phi}=1019\,\rm{MeV}$ \cite{PDG}. The mass difference between
the $\pi$ and  $\rho$ mesons are huge, we have to resort to large
hyperfine spin-spin interactions for explanation. The fine
spin-orbit  interactions and the hyperfine spin-spin interactions
are usually studied in the relativized quark model based on the
one-gluon exchange plus linear confinement potential motivated by
QCD \cite{Godfrey1985}, for more literatures, one can consult the
comprehensive  review Ref.\cite{Godfrey1985-RMP}. One may wander why
the contributions from the hyperfine interactions in Eq.(1) are so
large, and how to understand them in the quantum field theory.

At the energy scale $\mu=4\pi f_{\pi}\approx 1\,\rm{GeV}$, the
approximate chiral $SU_L(3)\times SU_R(3)$ symmetry  is
spontaneously broken to the $SU_V(3)$ symmetry by the small current
quark masses $m_u$, $m_d$ and $m_s$, and there appear eight
Nambu-Goldstone bosons (in the following we will neglect the word
Nambu for simplicity). The masses of the Goldstone bosons are
related with the current quark masses through  the
Gell-Mann-Oakes-Renner relation \cite{GMOR}. The quark fields $q(x)$
are usually decomposed as
\begin{eqnarray}
q(x)&=&\exp \left[-i\gamma_5\xi^a(x)\lambda^a\right]\widetilde{q}(x)
\, ,
\end{eqnarray}
where the $\widetilde{q}(x)$ and $\xi^a(x)$ denote the  constituent
quark fields (or Goldstone free fields) and the octet Goldstone
boson fields, respectively \cite{WeibergBook}, there exist
interactions among the quarks and  the Goldstone bosons. For
example, the spectra of the nucleons, $\Delta$ resonances and the
strange hyperons are well described by the constituent quark model
with the harmonic confinement potential plus the one-Goldstone-boson
exchanges induced potential \cite{One-goldstone}. In  the energy
region between the confinement and the spontaneous
 chiral symmetry breaking, the elementary degrees of freedom
are quarks, gluons and Goldstone bosons \cite{CH-Q-G}.

On the other hand, the mass breaking effects in the chiral doublets
are very large \cite{PDG,CH-2}, for example,
\begin{eqnarray}
\pi(0^-,140\,{\rm MeV}) &\leftrightarrow& f_0(600)(0^+,400-1200\,{\rm MeV}) \, ,\nonumber \\
\rho(1^-,775\,{\rm MeV}) &\leftrightarrow& a_1(1260)(1^+,1230\,{\rm MeV}) \, ,\nonumber \\
 p({\frac{1}{2}}^+,938\,{\rm MeV})&\leftrightarrow&N(1535)({\frac{1}{2}}^-,1525-1545\,{\rm MeV}) \, ,
\end{eqnarray}
which requires that the chiral symmetry should be badly broken, here
we use $\leftrightarrow$ to denote the chiral rotations. The chiral
massive quark dresses itself with the gluon cloud and
quark-antiquark pairs, and acquires a dynamically generated large
mass. We usually carry out the re-summation of the loops
nonperturbatively with the Dyson-Schwinger equation, and obtain the
Euclidean constituent quark masses by the definition $p^2=M^2(p^2)$,
which are compatible with the values used in the constituent quark
models \cite{Alkofer03,Roberts00}. The light pseudoscalar mesons
play a double role, as both Goldstone bosons and $q\bar{q}$ bound
states.

The exchanges of the one-gluon and one-Goldstone-boson between the
two  constituent quarks result in  the hyperfine interactions $H_C$
and $H_F$,  respectively \cite{One-goldstone,One-gluon},
\begin{eqnarray}
H_C &\sim& \frac{1}{M_iM_j}\vec{\lambda}_i^C \cdot \vec{\lambda}_j^C
\vec{\sigma}_i \cdot \vec{\sigma}_j \, , \nonumber \\
H_F &\sim& \frac{1}{M_iM_j}\vec{\lambda}_i^F \cdot \vec{\lambda}_j^F
\vec{\sigma}_i \cdot \vec{\sigma}_j \, ,
\end{eqnarray}
and they both contribute to the spin-spin interactions. In this
article, we take into account the contributions from the
one-Goldstone-boson exchange force besides the one-gluon exchange
force,   study the $\pi$ and $\rho$ mass difference with the
Bethe-Salpeter equation, and try to understand the difference
 in the quantum field theory.

   The  Bethe-Salpeter
equation is a conventional approach in dealing with the two-body
relativistic bound state problems \cite{BS51}, and has given many
successful descriptions of the hadron  properties in a Poincare
covariant way \cite{Roberts00,Roberts94}.

The article is arranged as follows:  we solve the Bethe-Salpeter
equation  for the $u\bar{d}$ bound  states  in Sec.2; in Sec.3, we
present the numerical results; and Sec.4 is reserved for our
conclusions.

\section{Bethe-Salpeter equation}
We write down the ladder Bethe-Salpeter equation for  the
  $u\bar{d}$ bound states in the Euclidean spacetime\footnote{In this article, we use the
metric $\delta_{\mu\nu}=(1,1,1,1)$, $\left\{\gamma_\mu
\gamma_\nu+\gamma_\nu \gamma_\mu\right\}=2\delta_{\mu\nu}$, the
momentums $k_\mu=(k_4,\overrightarrow{k})$,
$q_\mu=(q_4,\overrightarrow{q})$ and $P_\mu=(iE,\overrightarrow{P})$
with $P^2=-M^2_{\pi/\rho}$.},
\begin{eqnarray}
 S_{u}^{-1}\left(q+\xi_{u}P\right)\chi(q,P)S_{\bar{d}}^{-1}\left(q-\xi_{\bar{d}} P\right)&=&-\int
\frac{d^4
k}{(2\pi)^4}\left[\frac{\lambda^a}{2}\gamma_\mu \chi(k,P) \frac{\lambda^a}{2}\gamma_\mu \frac{g^2_s(q-k)}{(q-k)^2}\right.\nonumber\\
&&\left.+i\gamma_5 \chi(k,P) i\gamma_5 \frac{g^2_{\pi}(q-k)}{(q-k)^2+m^2_{\pi}}\right]\, , \\
 S_{u/\bar{d}}^{-1}\left(q\pm\xi_{u/\bar{d}}P\right)&=&i\left(\gamma \cdot q\pm \xi_{u/\bar{d}}
\gamma \cdot P\right)+M_{u/\bar{d}} \, ,\nonumber\\
\xi_{u/\bar{d}}&=&\frac{M_{u/\bar{d}}}{M_{u}+M_{\bar{d}}}\,
,\nonumber
\end{eqnarray}
 the $P_\mu$ is the four-momentum of the center of mass of the
$u\bar{d}$ bound states, the $q_\mu$ is the relative four-momentum
between the $u$ and $\bar{d}$ quarks, the $\chi(q,P)$ is the
Bethe-Salpeter amplitude of the $u\bar{d}$  bound states, and the
$g_\pi(q-k)$ and $g_s(q-k)$ are the energy dependent $\pi$-quark and
gluon-quark coupling constants,  respectively.
 In this article, we take the  $g^2(k)$ as a modified
Gaussian distribution, $g^2(k)=A
\left(\frac{k^2}{\Lambda^2}\right)^2\exp\left(-
\frac{k^2}{\Lambda^2}\right)$, where the strength  $A$ and the
distribution width $\Lambda^2$ are free parameters. The ultraviolet
behavior of the modified Gaussian distribution warrants that  the
integral in the Bethe-Salpeter equation is  convergent.

 The Euclidean Bethe-Salpeter amplitudes of the  $u\bar{d}$ bound states can be decomposed as
 \begin{eqnarray}
 \chi^\pi(q,P)&=&
 \gamma_5\left\{F_{\pi}(q,P)+i\!\not\!{P}F^\pi_1(q,P)
 +i\!\not\!{q} q \cdot P F^\pi_2(q,P)+\left[\!\not\!{P},\!\not\!{q}\right]F^\pi_3(q,P)
 \right\} \, , \nonumber\\
 \chi^\rho(q,P)&=&\!\not\!{\epsilon} \left\{
 iF_{\rho}(q,P)+\!\not\!{P}F^\rho_1(q,P)-\!\not\!{q}q \cdot PF^\rho_2(q,P)+i[\!\not\!{P},\!\not\!{q}]F^\rho_3(q,P)\right\}
 \nonumber\\
 &&+q \cdot \epsilon \left\{
 q \cdot PF^\rho_2(q,P)+2i \!\not\!{P}F^\rho_3(q,P) \right\} \nonumber\\
  &&+q \cdot \epsilon \left\{F^\rho_4(q,P)+i\!\not\!{P}q \cdot PF^\rho_5(q,P)-i\!\not\!{q}F^\rho_6(q,P)+[\!\not\!{P},\!\not\!{q}]F^\rho_7(q,P) \right\} \, ,
 \end{eqnarray}
 due to  Lorentz covariance \cite{Roberts94,VectorBS}, where the $\epsilon_\mu$ is the
 polarization vector of the $\rho$ meson, the $F_{\pi}(q,P)$, $F^\pi_i(q,P)$, $F_{\rho}(q,P)$ and $F^\rho_i(q,P)$ are the components
 of the Bethe-Salpeter amplitudes, which can be expanded  in terms of
Tchebychev polynomials $T^{\frac{1}{2}}_{n}(\cos \theta)$
\cite{Guth}, where  $\theta$ is  the included  angle between $q_\mu$
and $P_\mu$. Numerical calculations indicate that taking only the
terms $T^{\frac{1}{2}}_{0}(\cos \theta)=1$  can give satisfactory
results \cite{WangNPA-PRC}. If we take into account the small terms
with $n\geq 1$, the predictions may be improved mildly.  In the
following, we use the amplitudes  $ F_{\pi/\rho}(q^2,P^2)$ and $
F_i^{\pi/\rho}(q^2,P^2)$ to denote the $n=0$ terms of the
Bethe-Salpeter amplitudes $F_{\pi/\rho}(q,P)$ and
$F_i^{\pi/\rho}(q,P)$, respectively. Then the Bethe-Salpeter
equations can be projected into four and eight coupled integral
equations for the $\pi$ and $\rho$ mesons respectively, and it is
very difficult to solve  them numerically. Furthermore, we cannot
obtain physical insight from those  involved integral equations.

 Multiplying both sides of the Bethe-Salpeter equations  of the $\pi$ and $\rho$
mesons by
 $\gamma_5\left[\!\not\!{q},\!\not\!{P}\right]$ \cite{WangCPL},
  and
 $\!\not\!{\epsilon}\left[\!\not\!{q},\!\not\!{P}\right]+\left[\!\not\!{q},\!\not\!{P}\right]\!\not\!{\epsilon}$,
 $q \cdot \epsilon q \cdot P \!\not\!{P}$ respectively,  completing the trace
 in the Dirac spinor space, carrying out the integrals for the included angle  $\theta$,
 and neglecting the small components  $F^\pi_3$, $F^\rho_3$, $F^\rho_5$ and
  $F^\rho_6$, we can obtain the following three
 relations,
 \begin{eqnarray}
 F_{\pi}(q^2,P^2)-\left(M_u+M_{\bar{d}}\right)F_1^\pi(q^2,P^2)&=&0 \, , \nonumber \\
 F_{\rho}(q^2,P^2)+\left(M_u+M_{\bar{d}}\right)F_1^\rho(q^2,P^2)&=&0 \, , \nonumber \\
 2F_{\rho}(q^2,P^2)-\left(M_u+M_{\bar{d}}\right)F_4^\rho(q^2,P^2)&=&0 \, .
 \end{eqnarray}

 The Bethe-Salpeter  amplitudes can be approximated   as
\begin{eqnarray}
 \chi^\pi(q,P)&=&
 \gamma_5\left(1+\frac{i\!\not\!{P}}{M_u+M_{\bar{d}}}\right)F_{\pi}(q^2,P^2)  \, ,\nonumber\\
 \chi^\rho(q,P)&=&\left\{\!\not\!{\epsilon} \left(i-\frac{\!\not\!{P}}{M_u+M_{\bar{d}}}\right)+ \frac{2q \cdot \epsilon}{M_u+M_{\bar{d}}} \right\}F_{\rho}(q^2,P^2)
 \, ,
 \end{eqnarray}
if we also neglect the small components $F^\pi_2$,  $F^\rho_2$ and
$F^\rho_7$. Then the involved Bethe-Salpeter equations can be
projected into the following simple form,
 \begin{eqnarray}
  \left\{ q^2+M_{u}M_{\bar{d}}\left[1+\frac{P^2}{\left(M_{u}+M_{\bar{d}}\right)^2}\right]\right\}F_{\pi}(q^2,P^2)&=&\int \frac{d^4k}{(2\pi)^4} F_{\pi}(k^2,P^2)  \nonumber\\
  &&\left\{\frac{16}{3}\frac{g^2_s(q-k)}{(q-k)^2} +\frac{g^2_\pi(q-k)}{(q-k)^2+m_\pi^2} \right\}\, , \nonumber\\
  \left\{q^2+M_{u}M_{\bar{d}}\left[1+\frac{P^2}{\left(M_{u}+M_{\bar{d}}\right)^2}\right]\right\}F_{\rho}(q^2,P^2)&=&\int\frac{d^4k}{(2\pi)^4}F_{\rho}(k^2,P^2)  \nonumber\\
  &&\left\{\frac{8}{3}\frac{g^2_s(q-k)}{(q-k)^2} -\frac{g^2_\pi(q-k)}{(q-k)^2+m_\pi^2} \right\}\,  .
 \end{eqnarray}

 If we take $q^2=0$ and $g^2_\pi=0$, and assume  that there exists a
physical solution, then
\begin{eqnarray}
 M_{u}M_{\bar{d}}\left[1-\frac{M_{\pi/\rho}^2}{(M_{u}+M_{\bar{d}})^2}\right]F_{\pi/\rho}(0,-M_{\pi/\rho}^2)&=&\int \frac{d^4k}{(2\pi)^4} F_{\pi/\rho}(k^2,-M_{\pi/\rho}^2)G_{\pi/\rho}(0-k) \,
 ,\nonumber\\
\end{eqnarray}
where the $G_{\pi/\rho}(k)$ denotes the interacting kernels.  In
numerical calculations, we observe that the Bethe-Salpeter amplitude
$F_{\pi/\rho}(k^2,-M_{\pi/\rho}^2)$ has the same sign in the region
$k^2\geq0$,
\begin{eqnarray}
 1-\frac{M_{\pi/\rho}^2}{(M_{u}+M_{\bar{d}})^2}&=&\int\frac{d^4k}{(2\pi)^4}\frac{F_{\pi/\rho}(k^2,-M_{\pi/\rho}^2)}{M_{u}M_{\bar{d}}F_{\pi/\rho}(0,-M_{\pi/\rho}^2)}G_{\pi/\rho}(0-k) >0\, ,
\end{eqnarray}
 and obtain an simple relation (or constraint),
\begin{eqnarray}
M_{\pi/\rho}^2<(M_{u}+M_{\bar{d}})^2 \, ,
\end{eqnarray}
which survives for $q^2>0$ (although the relation is not explicit
for $q^2>0$), i.e. the bound energy $E_{\pi/\rho}$ originates from
the interacting kernel $G_{\pi/\rho}(k)$ and should be negative,
$E_{\pi/\rho}=M_{\pi/\rho}-M_{u}-M_{\bar{d}}<0$. The numerical
calculations indicate that above arguments survive  in the case
$g_{\pi}^2\neq 0$.

From Eq.(9),  we can see explicitly that the one-gluon exchange
force in the $\pi$ channel is more attractive than that in the
$\rho$ channel due to the factors $\frac{16}{3}$ and $\frac{8}{3}$;
furthermore, the one-pion exchange force is attractive in the $\pi$
channel and repulsive in the $\rho$ channel,  the bound energy
$|E_\pi|\gg |E_\rho|$ can be accounted  for naturally.

We can introduce a parameter $\lambda(P^2)$ and solve  above
equations as an eigenvalue problem, the masses of the $\pi$ and
$\rho$ mesons can be determined by the condition
$\lambda(P^2=-M_{\pi/\rho}^2)=1$,
\begin{eqnarray}
  \left\{ q^2+M_{u}M_{\bar{d}}\left[1+\frac{P^2}{\left(M_{u}+M_{\bar{d}}\right)^2}\right]\right\}F_{\pi}(q^2,P^2)&=&\lambda(P^2)\int \frac{d^4k}{(2\pi)^4} F_{\pi}(k^2,P^2)  \nonumber\\
  &&\left\{\frac{16}{3}\frac{g^2_s(q-k)}{(q-k)^2} +\frac{g^2_\pi(q-k)}{(q-k)^2+m_\pi^2} \right\}\, , \nonumber\\
  \left\{q^2+M_{u}M_{\bar{d}}\left[1+\frac{P^2}{\left(M_{u}+M_{\bar{d}}\right)^2}\right]\right\}F_{\rho}(q^2,P^2)&=&\lambda(P^2)\int\frac{d^4k}{(2\pi)^4}F_{\rho}(k^2,P^2)  \nonumber\\
  &&\left\{\frac{8}{3}\frac{g^2_s(q-k)}{(q-k)^2} -\frac{g^2_\pi(q-k)}{(q-k)^2+m_\pi^2} \right\}\, . \nonumber\\
 \end{eqnarray}

\section{Numerical results}
 The constituent quark masses of the $u$ and $\bar{d}$ quarks are taken as
 $M_u=M_{\bar{d}}=400\,\rm{MeV}$.
 The strength parameter $A$ and the distribution width $\Lambda$ are
 free parameters, we take the values $\Lambda=200\,\rm{MeV}$ and
 $A=146 (105)$  for the one-gluon (one-pion) exchange. Other values
 of the $\Lambda$ and $A$ also work, we choose the present
 parameters for illustration.

We solve the Bethe-Salpeter equations  as  an eigenvalue  problem
numerically by direct iterations, and observe the convergent
behaviors  are  very good.
    The numerical results for the Bethe-Salpeter amplitudes
are shown in Fig.1. From the figure, we can see that   the
Bethe-Salpeter amplitudes center around zero momentum and extend to
the energy scale about $q=0.4\,\rm{GeV}$ and $0.7\,\rm{GeV}$ for the
$\rho$ and $\pi$ mesons respectively, the stronger interactions in
the $\pi$ channel result in more stable bound state   than that in
the $\rho$ channel, as the bound energies are $E_\pi=-660\,\rm{MeV}$
and $E_{\rho}=-25\,\rm{MeV}$, respectively. In numerical
calculations, we observe that the one-pion exchange force plays an
important role and should be taken into account.

With the following simple replacements in Eq.(13),
\begin{eqnarray}
M_{\bar{d}} &\to&M_{\bar{s}}=536\,{\rm{MeV}} \, ,\nonumber \\
g^2_{\pi}(k^2)&\to&g^2_{K}(k^2)=A\left(\frac{k^2}{\widetilde{\Lambda}^2}\right)^2\exp\left(-\frac{k^2}{\widetilde{\Lambda}^2}\right)\, ,\nonumber \\
M_{\pi}&\to&M_{K}=494\,\rm{MeV} \, ,
\end{eqnarray}
where the $\widetilde{\Lambda}=\Lambda
\left(\frac{M_K+M_{K^*}}{M_{\pi}+M_{\rho}}\right)^2$ denotes the
$SU_V(3)$ breaking effect for the coupling constant, we can obtain
the corresponding solutions for the pseudoscalar meson $K$ and
vector meson $K^*$ with the eigenvalues
$\lambda(P^2=-M_K^2=-(494\,\rm{MeV})^2)=1$ and
$\lambda(P^2=-M_{K^*}^2=-(892\,\rm{MeV})^2)=1$, respectively. The
bound energies are $E_{K}=-442\,\rm{MeV}$ and
$E_{K^*}=-44\,\rm{MeV}$, respectively, which indicate that there
exists  a more stable bound state in the pseudoscalar channel than
that in the vector channel.

\begin{figure}
 \centering
 \includegraphics[totalheight=7cm,width=8cm]{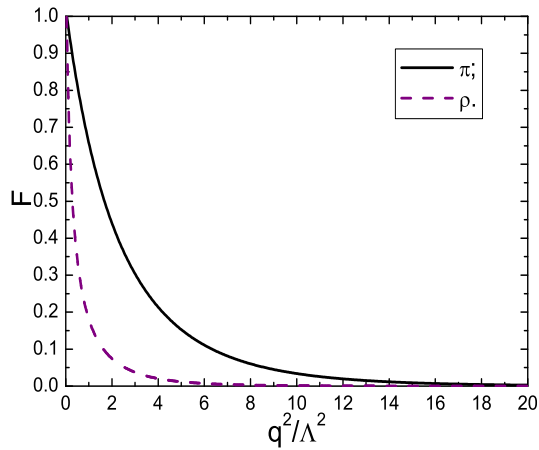}
  \caption{The Bethe-Salpeter amplitudes of the bound states.  }
\end{figure}

\section{Conclusion}
In this article, we take  into account the one-pion exchange force
besides   the one-gluon exchange force to study the mass difference
of the $\pi$ and $\rho$ mesons with the Bethe-Salpeter equation.
After simplifying the involved Bethe-Salpeter equations, we observe
that the one-gluon exchange force in the $\pi$ channel is more
attractive than that in the $\rho$ channel, while the one-pion
exchange  force is attractive in the $\pi$ channel and repulsive in
the $\rho$ channel,
  the bound energy $|E_\pi|\gg |E_\rho|$ can be accounted  for naturally.

\section*{Acknowledgments}
This  work is supported by National Natural Science Foundation,
Grant Number 11075053, and Program for New Century Excellent Talents
in University, Grant Number NCET-07-0282, and the Fundamental
Research Funds for the Central Universities.

\end{document}